\pdfoutput=1 
\documentclass{JINST}

\usepackage{amsmath}

\title{Neutron detectors for the ESS diffractometers}

\author{I. Stefanescu$^a$\thanks{Corresponding author.}, M. Christensen$^b$, J. Fenske$^c$, R. Hall-Wilton$^{a,d}$, P. F. Henry$^{a}$, O. Kirstein$^{a,e}$, M. M\"uller$^c$, G. Nowak$^c$,
D. Pooley$^f$, D. Raspino$^f$, N. Rhodes$^f$, J. \v Saroun$^g$, J. Schefer$^h$, E. Schooneveld$^f$, J. Sykora$^f$, W. Schweika$^{a,i}$\\
\llap{$^a$}European Spallation Source ESS ERIC, P.O. Box 176, SE-221 00, Lund, Sweden\\
\llap{$^b$}Center for Materials Crystallography  iNANO$\&$Department of Chemistry, Aarhus University., DK-8000 Aarhus, Denmark\\
\llap{$^c$}Helmholtz-Zentrum Geesthacht, Max-Planck-Stra\ss e 1, 21502 Geesthacht, Germany\\
\llap{$^d$}Mid-Sweden University, SE-85170 Sundsvall, Sweden\\
\llap{$^e$}University of Newcastle, NSW 2308, Australia\\
\llap{$^f$}ISIS Department, Science and Technology Facilities Council, Rutherford Appleton Laboratory, Harwell Oxford, Oxfordshire, OX11 0QX, U.K\\
\llap{$^g$}Nuclear Physics Institute, Husinec-\v Re\v z 130, 25068 \v Re\v z, Czech Republic\\
\llap{$^h$}Laboratory for Neutron Scattering and Imaging, Paul Scherrer Institut, CH-5232 Villigen PSI, Switzerland\\
\llap{$^i$}Forschungszentrum J\"ulich GmbH, 52425 J\"ulich, Germany\\
E-mail: \email{irina.stefanescu@esss.se}}

\abstract{The ambitious instrument suite for the future European Spallation Source whose civil construction started recently in Lund, Sweden, demands a set of diverse and challenging requirements for the neutron detectors. For instance, the unprecedented high flux expected on the samples to be investigated in neutron diffraction or reflectometry experiments requires detectors that can handle high counting rates, while the investigation of sub-millimeter protein crystals will only be possible with large-area detectors that can achieve a position resolution  as low as 200 $\mu$m. This has motivated an extensive research and development campaign to advance the state-of-the-art detector and to find new technologies that can reach maturity by the time the ESS will operate at full potential.  This paper presents the key detector requirements for three of the Time-of-Flight (TOF) diffraction instrument concepts selected by the Scientific Advisory Committee to advance into the phase of preliminary engineering design. We discuss the  detector technologies commonly employed at the existing similar instruments and their major challenges for ESS. The detector technologies selected  by the instrument teams to collect the diffraction patterns are also presented.  Analytical calculations, Monte-Carlo simulations, and real experimental data are used to develop a generic method to estimate the event rate in the diffraction detectors.  We apply this method to make predictions for the future diffraction instruments, and thus provide additional information that can help the instrument teams with the optimisation of the detector designs. }


\begin{document}

\section{Introduction}\label{sec:intro}

The European Spallation Source (ESS) currently under construction in Lund, Sweden, is expected to attain a peak neutron brightness of $\sim$ 8$\cdot$10$^{14}$ n$\cdot$cm$^{-2}\cdot$s$^{-1}\cdot$sr$^{-1}$$\cdot$\AA$^{-1}$ for both the cold and thermal spectra, with a peak proton current of  62.5 mA and a pulse length of 2.86 ms at a repetition rate of 14 Hz ~\cite{ess}.  This facility will be part of the future suite of European research infrastructures that will provide experimental opportunities for research with neutrons for both academia and industry. The first neutrons from the source are expected in 2019, and the facility will reach its full design specifications in 2025.  A baseline suite of 22 neutron scattering instruments will be available to users when the facility is completed. These instruments  will be using the intense neutron beam to study the structure and dynamics in materials by using a broad spectrum of techniques including diffraction, reflectometry, imaging, quasielastic and inelastic scattering \cite{tdr}.  

As neutron diffraction has a very broad range of scientific applications and a large user community at all existing neutron scattering facilities, powder diffractometers are a leading priority for the new instrumentation at ESS. This research field has made tremendous progress with the advent of bright spallation sources, such as Target Station 2 at ISIS, UK \cite{isis}, SNS in the USA \cite{sns} and J-PARC in Japan \cite{jparc}. Also, the continuous upgrade of the instrumentation at the world's most intense reactor source ILL in France \cite{ill}, ensured that the powder diffractometers operating there have one of the highest scientific output worldwide  \cite{esfri}. These facilities pushed forward the development of cutting edge instrumentation that includes complex sample environments, advanced guide and chopper systems,  highly-efficient 2D-detector systems, and sophisticated data analysis.  Some of them already operate suites of  diffractometers covering hot, thermal and cold neutron ranges \cite{wish1,powgen0,superhrpd}, and that can perform measurements at extreme conditions of temperature and pressure \cite{snap,nova}.  Investigation of structure and dynamics of materials can be achieved by using multiple techniques by combining neutron diffraction with either Small Angle Neutron Scattering (SANS) \cite{spica}  or imaging \cite{imat}  in the same instrument.  

A balanced suite of diffraction instruments at the ESS must follow this well-established path, requiring several instruments optimised for parts of the science case that offer complementary capabilities and measurement types. The long-pulse structure unique to the ESS beam and the unprecedented source brightness provide the flexibility to tailor the instrument  Q-range\footnote[1]{Q is the magnitude of the scattering vector (transferred momentum) for elastic scattering i.e., assuming no change of neutron energy in the scattering process. It is defined as $Q=(4\pi/\lambda)\cdot\sin\theta$, where $\lambda$ is the neutron wavelength and $\theta$ is the scattering angle.}  and resolution as desired, and enable investigations of multiple dimensions at fast time scales  \cite{tdr}. Also, the high flux expected to be available at the sample position will make it possible to study samples with volumes below 1 mm$^3$. In order to meet these challenging scientific goals, several new hardware components, such as chopper systems, beam optics and  detectors, as well as  sophisticated data analysis and data handling methods, must be developed. The new instruments must be supported by well equipped sample preparation, handling and characterisation laboratories.  The detection systems must allow for a very fast collection of the diffraction patterns over a wide angular range, and this information must be fully integrated  with the signals provided by the accelerator, sample environments, chopper control systems, etc. The future users will need massive computational resources and fast local access to data \cite{dmsc}.  Fortunately, significant amounts of expertise in all these areas are already in place at the existing neutron research centres. As the development of the ESS instrument suite is a joint European venture, the experience of the partner institutes is well matched to the demands of this project. The ESS will not only capitalise and build upon the vast body of existing expertise, but also establish links to similar efforts throughout Europe \cite{tdr}. 

The challenging requirements for the detectors to be deployed at the high-intensity  neutron sources were recognised by the scientific community already in the early 2000s. With several new ILL instruments coming online or needing upgrades, the SNS and JPARC spallation sources under construction, and the planning for the long-pulse spallation facility ESS advancing rapidly,  it became clear that there will be a growing demand for technologies that can provide highly segmented detectors with a high count rate capability. Conferences and  workshops organised regularly by the detector community aimed at identifying the common detector needs across these facilities, sharing  knowledge and experience, and presenting both low-risk and highly innovative ideas for future developments \cite{conf1,conf2,conf3,conf4}. These meetings also served as a basis for the  establishment of international collaborations, working groups or joint activities in order to efficiently coordinate the effort to advance the technological development of performant neutron detectors, such as the International Collaboration for Neutron Detectors (ICND), which includes the detector experts representing all major neutron scattering facilities \cite{icnd,karl}. Several of the emerging detector and readout concepts received funding from the European Commission \cite{nmi3,detni,crisp}, or from the national funding agencies \cite{powgen,Hen12,d1b,gem}. Most of these developments led to detectors that  are now operational at ILL or elsewhere \cite{bruno}.  For example, the MILAND Multi Wire Proportional Counter (MWPC), deployed since 2008 at the D16 instrument at ILL \cite{d16}, was developed within a joint research activity funded by an EU grant \cite{miland}. MILAND is a $^3$He-based detector with a sensitive area of 32 cm $\times$ 32 cm and 1 mm readout pitch. The MILAND detector can achieve a global countrate of 0.7 MHz with 10$\%$ deadtime, which makes it one of the fastest MWPCs used for neutron detection.  Another development that  led to several operational detectors for use at high-intensity neutron sources is the  Microstrip Gas Chamber (MSGC) \cite{clergeau}, which exploits the micro-strip technology \cite{oed}. MSGC-based detectors are  installed today at the D4 and D20 instruments \cite{msgc,d20} at ILL. The D20 detector can reach a count rate of 20 kHz per 3.2$^\circ$-cell at 10$\%$ deadtime and provide an intrinsic resolution of 3 mm \cite{d20}. 

The state-of-the-art  detection systems  deployed at diffractometers recently commissioned at the existing spallation sources consist of hundreds of $^3$He-filled position-sensitive tubes or scintillator modules, the only two technologies able to cover several square meters of sensitive areas as  required by this specific instrument class, and also fulfill the requirements for efficiency and position resolution.  
For the scintillator-based diffraction detectors, the neutron converter could be either ZnS:$^6$LiF(Ag)/ZnS:$^{10}$B$_2$O$_3$ screens  \cite{powgen,powgen11,gem,gem1,gem2,senju,ibix} or a Li-glass \cite{topaz} coupled to a photosensor via wavelength shifting or clear fibers.  With this technology, pixel sizes as small as 1 mm$^2$ became possible due to the advances made in the commercial production of high-efficiency, low dark-rate multi-anode photomultiplier tubes (PMTs) or silicon photo-multipliers (SiPMs) \cite{ibix,poldi}. The efficiency of a scintillator-based detector can now reach 65$\%$ at 1.2 \AA~\cite{poldi3}. The maximum count rate capability of  a scintillator module depends upon the type and thickness of the scintillator material, as well as the fiber-photosensor coupling or encoding scheme.  For example, the maximum count rate reported for the ISIS linear position sensitive wavelength shifting fiber (WLSF) detector is 16 kHz per PMT \cite{syk12}. This detector consists of a set of 16 pixels, each having a width of 2 mm  and a length of 200 mm. The pixels were coded such that four scintillator elements were viewed by a single PMT \cite{syk12}.  

The intense efforts carried out over the last years into increasing the performance of scintillator-based detectors  in terms of efficiency and counting rate capability showed positive results from the perspective of operation under the intense neutron fluxes expected at spallation neutron sources. The combined neutron imaging and neutron diffraction beamline IMAT, currently under construction at the ISIS facility \cite{imat}, and the POLDI  diffractometer operational at the Swiss spallation source SINQ at PSI  \cite{poldi,poldi3,poldi2}, are examples of instruments that will be populated with scintillator modules combining the latest  achievements in the field of photosensors or signal readout.  

WISH (ISIS),  NOMAD (SNS), TAIKAN and SuperHRPD (J-PARC)  are examples of diffractometers operated with hundreds of $^3$He-filled position-sensitive tubes surrounding the sample \cite{wish1,superhrpd,wish2,nomad,taikan}.  Tubes with diameters of 25, 12.5 or 8 mm and lengths of 1 m or less are mounted in panels of eight or ten, which  can be arranged in several ways with respect to the scattering plane. The position resolution in one dimension is given by the tube diameter, which for large area detectors of several square metres, is currently limited to 8 mm (e.g. WISH@ISIS). Finer granularity will be harder to reach, as it requires further reduction of the tube diameter and increase of the gas pressure.  The efficient operation of an 8-mm diameter tube requires between 10 and 15 bar of $^3$He gas and the reported maximum count rate  capability is around 150 kHz \cite{wish1,wish2}.  

In this paper we give a brief description of the key design specifications for the main features of the ESS powder and materials science diffraction  instruments. A detailed description  of the instrument design and scientific goals can be found  in the instrument proposals available on the ESS public website \cite{dream,heimdal,beer}.  We  focus here on the envisaged diffraction detectors and their  requirements. We present  the strategy for the diffraction detectors at the ESS and discuss briefly the detector solutions proposed by the respective instrument teams.  It is clear that in order to fully exploit the intense ESS beam and also satisfy the scientific demands  of this particular instrument class, the selected detector technologies must be able to cover large areas, provide the required position resolution and have high count rate  capabilities. Furthermore, we propose  an analytical method to determine the rates in the detectors used to collect the neutron diffraction patterns, which is based on knowledge obtained from real experimental data. We apply this method to estimate the event rates in the future ESS diffraction detectors and compare these rates  to those obtained from  the Monte-Carlo simulations performed by the instrument teams in order to optimise the instrument designs.

\section{Diffraction instruments for the ESS}

The instrument suite presented in the ESS Technical Design Report published in 2013 includes five diffractometers \cite{tdr}. The Scientific Advisory Committee in charge of the selection process during the 2013-2014 proposal round endorsed the following instruments:

\begin{itemize}
\item DREAM ({\bf D}iffraction {\bf R}esolved by {\bf E}nergy and {\bf A}ngle {\bf M}easurement), a bi-spectral powder diffractometer aimed at studying large unit cells and magnetic structures. This instrument will be built by a collaboration consisting of scientists from the J\"ulich Centre for Neutron Science at the Research Centre J\"ulich, Germany, and the Laboratoire Leon Brillouin at the CEA Saclay, France \cite{dream,dream1}.  
\item Heimdal, a thermal powder diffractometer with added capabilities for small-angle neutron scattering (SANS) and neutron imaging (NI) dedicated to the investigation of functional materials. It  was proposed by a collaboration involving scientists from the Aarhus University, the Niels Bohr Institute and the Technical University of Denmark in Denmark,  the Institute for Energy Technology Oslo, Norway, and the Paul-Scherrer Institute, Switzerland \cite{heimdal,heimdal_nima}. 
\item BEER ({\bf B}eamline for {\bf E}uropean materials {\bf E}ngineering {\bf R}esearch), a material science and engineering diffractometer for use by the applied research community. Like Heimdal, BEER is also designed to include dedicated SANS and NI setups, which will provide more insight into the structure of the sample under investigation.  This instrument will be built by a collaboration between the Helmholtz-Zentrum Geesthacht, Germany, and the Nuclear Physics Institute in  the Czech Republic \cite{beer,beer1}. 

\end{itemize}

The SANS and NI options for Heimdal and BEER are foreseen as upgrades to the initial powder diffraction set-ups and require the installation of vacuum tanks hosting the dedicated SANS detectors, as well as the neutron imaging systems. This will increase the complexity of the instruments and impose design constraints in terms of geometry and size, but it will enable new types of experiments to be performed at the same beamline.  

All three diffractometers will use a substantial part of the full white beam delivered by the ESS source. Their bandwidth, denoted here as $\Delta\lambda$, will be limited by their length and choice of pulse suppression \cite{dream,dream1,heimdal,heimdal_nima,beer,beer1}. Pulse-shaping choppers will be used to provide a sharp time structure that will define the instrumental resolution.  This lends a high degree of versatility to the instruments, which  can operate in either high resolution (e.g., $\Delta d/d\footnote[1]{$d$ is the distance between the lattice planes of a crystal. It is defined by the Bragg equation $d=n\lambda/(2\cdot\sin\theta)$ , where $n$ is the order of the Bragg reflection, $\theta$ is the scattering angle and $\lambda$ is the wavelength of the neutron.}\sim$2$\cdot$10$^{-4}$ \cite{dream}) or high intensity mode. Results of Monte-Carlo simulations of the beamlines indicate that by degrading the instrument resolution to 1$\%$, the flux on the sample can be increased  by almost two orders of magnitude when compared to the high resolution mode of operation \cite{dream,heimdal}. For example, the estimated time-averaged flux over the wavelength interval 0.6-2.3 \AA~at the Heimdal sample position for $\Delta d/d$=1$\%$ at 90$^\circ$ can be as high as 10$^9$ n/s/cm$^2$. This is more than one order of magnitude greater than the highest flux achieved on the POWGEN or WISH samples in the same experimental conditions \cite{dream}. It is important to mention here that the flux calculations included in all instrument proposals submitted to the 2013 evaluation round were made for the geometry of the coupled moderator presented in the TDR \cite{tdr}. The new optimised thermal moderator design developed recently is expected to provide an increase of up to 35$\%$ in the flux at the sample position for all baseline diffraction instruments \cite{Mezei2014}. 

The approved diffraction instruments are scheduled to enter the preliminary engineering design phase in 2016. The detector designs foresee complex systems that require advanced engineering and must share the narrow space available around the sample with complicated sample environments, radial collimators, and shielding. Powder diffractometers should be operational when the first neutrons become available from the source and therefore the diffraction detectors may be prioritised during the preliminary design phase of the instruments.

\subsection{DREAM, the ESS bi-spectral powder diffractometer}

The bi-spectral powder diffractometer DREAM will be specialised in the study of magnetic systems and for materials with large unit cells. The advantage of the new ESS diffractometer over the existing similar instruments will be the expanded spectral range that will be available at the sample position, as well as the high flexibility for trading resolution versus intensity through appropriate pulse-shaping.  DREAM will use frame multiplication to include both the thermal and cold neutrons in the collection of the powder diffraction patterns.  It will operate with a wavelength band of $\Delta \lambda$=3.6 \AA ~using all the ESS pulses. Pulse-shaping to 10 $\mu$s will yield an ultimate high resolution of 0.03$\%$ at $\lambda$=1.5 \AA. The neutron optics will be optimised for a  vertical and horizontal divergence at the sample position of 0.5$^\circ$  \cite{dream}.     

\begin{figure*}[ht]
\centering
\includegraphics[scale=0.35]{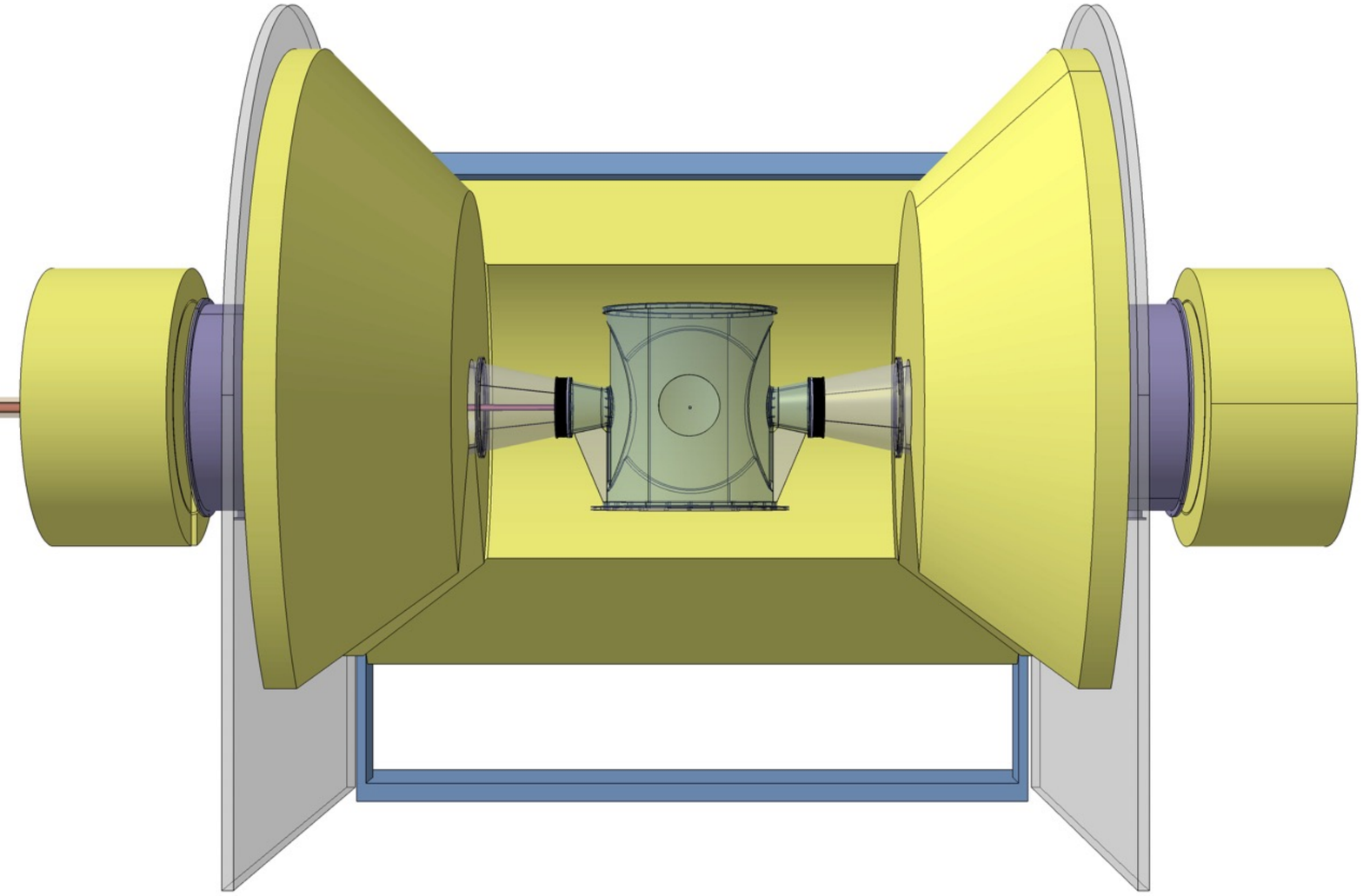}
\caption{Cartoon showing the DREAM sample area with the cylindrical detector covering 6.2 sr,  forward and backward detectors. }  
\label{fig: f1}
\end{figure*}

The sample position will be at 76.5 m from the moderator and it will be surrounded by a detector with cylindrical geometry, with the axis common to the beam axis and a radius of 1.25 m, see Fig. \ref{fig: f1}. The solid angle covered by the detector will be close to 6.2 sr. The background will be reduced by evacuating the primary flight path and filling the sample tank with argon gas at atmospheric pressure. The required detector spatial resolution is 4 mm $\times$ 4 mm, which is matched to the incoming symmetric beam divergence.  The high spatial resolution of the detector will enable additional single-crystal measurement capabilities. 

This instrument will also host a 1 m$^2$ back-scattering detector covering the section between 168 and 177$^\circ$, which will be used to study diffraction patterns with very high resolution. A similar detector is foreseen for the forward direction  in order to include the region 0.01 \AA$^{-1} <$Q$<$1 \AA$^{-1}$,  important when investigating powder diffraction of nano-structured materials.

\subsection{Heimdal, the ESS thermal powder diffractometer}

Heimdal will be the ESS thermal powder diffractometer  dedicated to the {\it in situ} and {\it in operandi} study of the functional materials important for chemistry, crystallography, physics, and materials science. It is proposed to have a length of 167 m and will combine state-of-the-art thermal neutron powder diffraction, small-angle neutron scattering (SANS) and neutron imaging (NI). Its design is based on a novel concept, where two independent guides view the cold and thermal moderators, respectively, and extract and transport the beams to the sample position. The thermal guide will be optimised for powder diffraction studies, while the beam delivered by the cold guide will be used to perform small-angle scattering and neutron imaging measurements \cite{heimdal,heimdal_nima}.  

\begin{figure*}[ht]
\centering
\includegraphics[scale=0.55]{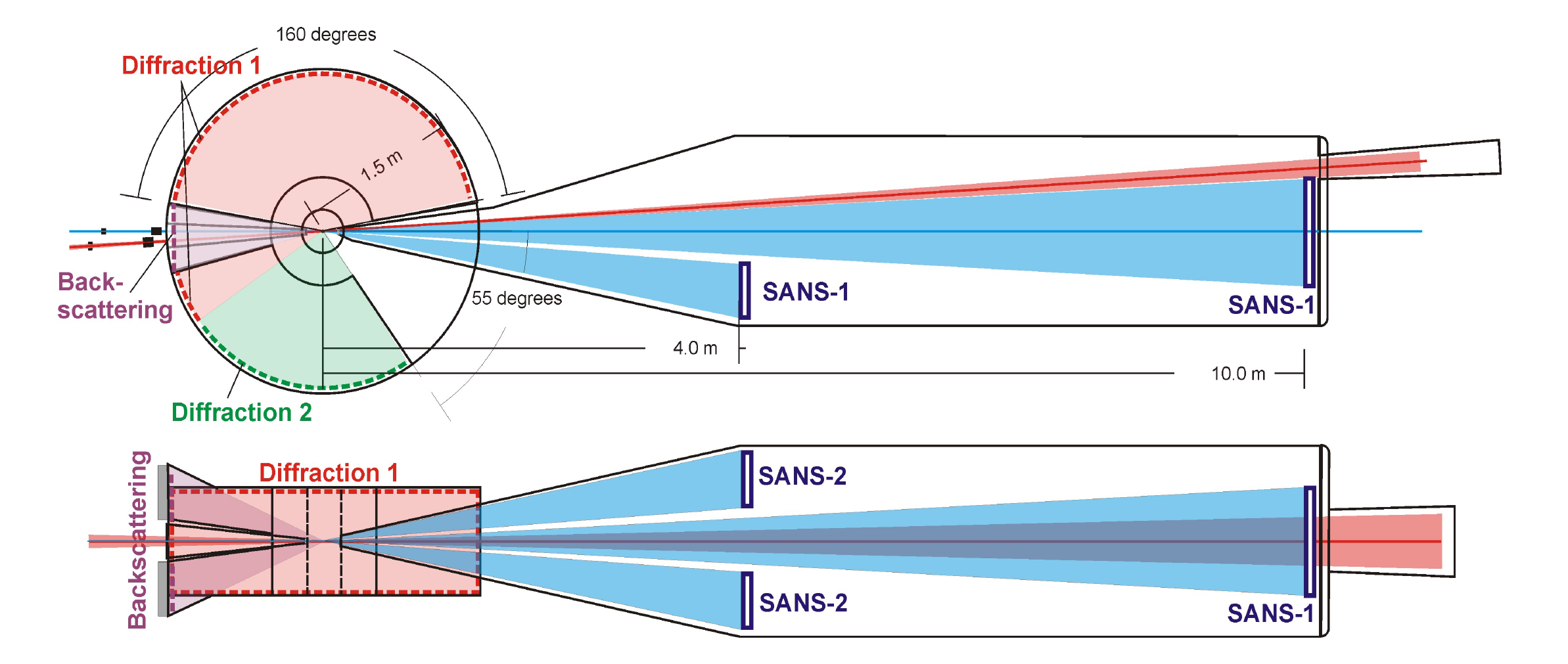}
\caption{Technical drawing of the Heimdal sample area showing the powder diffraction detectors surrounding the sample, the SANS tank with the detectors inside, and the back-scattering detectors above and below the incoming beam. The detectors labeled as ``Diffraction 1`` and shown in red are the Day-1 detectors and those labeled as ``Diffraction2`` and shown in green will be added later. Taken from  \cite{heimdal}.}  
\label{fig: f2}
\end{figure*}

The Heimdal instrument will cover diffraction patterns in the $Q$-range from 0.6-21 \AA$^{-1}$ by using a cylindrical detector arrangement covering an angle from 10-170$^\circ$ on one side, and 170-150$^\circ$ on the other side, see Fig. \ref{fig: f2}. The area of the powder diffraction detectors will be close to 4 m$^2$. The vertical coverage of these detectors will be limited to 1 m ($\pm$18$^\circ$). The pixel resolution of 3$\times$10 mm$^2$ was chosen to match a sample size of 5$\times$15 mm$^2$. The volume between the sample and detectors between 0.5 and 1.5 m will be filled with Ar gas or dry air in order to reduce scattering of the neutrons. 

Back-scattering detectors mounted above and below the beam will provide additional data for high-$Q$-values, see Fig. \ref{fig: f2}. The green area labeled as ``Diffraction 2``  will also be filled with diffraction detectors, but at a later stage.  The same figure shows the SANS tank that will host four dedicated SANS detectors. The first one will be located 10 m downstream of the sample and consist of a flat panel with a size of 1 m $\times$ 1m and a pixel resolution of 4 mm $\times$ 4 mm.  The other three SANS detectors will have a size of 0.5 m$^2$ each and be placed at 4 m from the sample,  displaced with respect to the central beam, see Fig. \ref{fig: f2}.  The neutron imaging station will be located inside the sample chamber, and will be operated whenever needed, but not during SANS measurements.    

\subsection{BEER, the ESS materials science and engineering diffractometer}

BEER will be the ESS engineering diffractometer dedicated to {\it in situ} studies of the industrial processing of materials, residual stress mapping and crystallographic texture \cite{beer,beer1}. The sample environment will include an advanced thermomechanical physical simulator for material processing (Gleeble simulator) \cite{gleeble}. This instrument will also allow for the possibility of studying simultaneously various aspects of the microstructure of materials, e.g., phase content, strain or texture of nano-particles by combining powder diffraction with SANS or NI. 

The BEER instrument will be 159 m long, and its resolution can be easily defined with the help of pulse-shaping and multiplexing choppers.  A  wavelength range including thermal and cold neutrons with $\lambda$ between 0.5 and 7.5 \AA~can be selected by using the method of alternating wavelength frames from subsequent pulses.  The lower wavelength range (1.2-2.9 \AA) will be used for powder diffraction analysis, while the colder part of the spectrum (4.7-6.3 \AA) is appropriate for performing simultaneous  SANS measurements.     

\begin{figure*}[ht]
\centering
\includegraphics[scale=0.55]{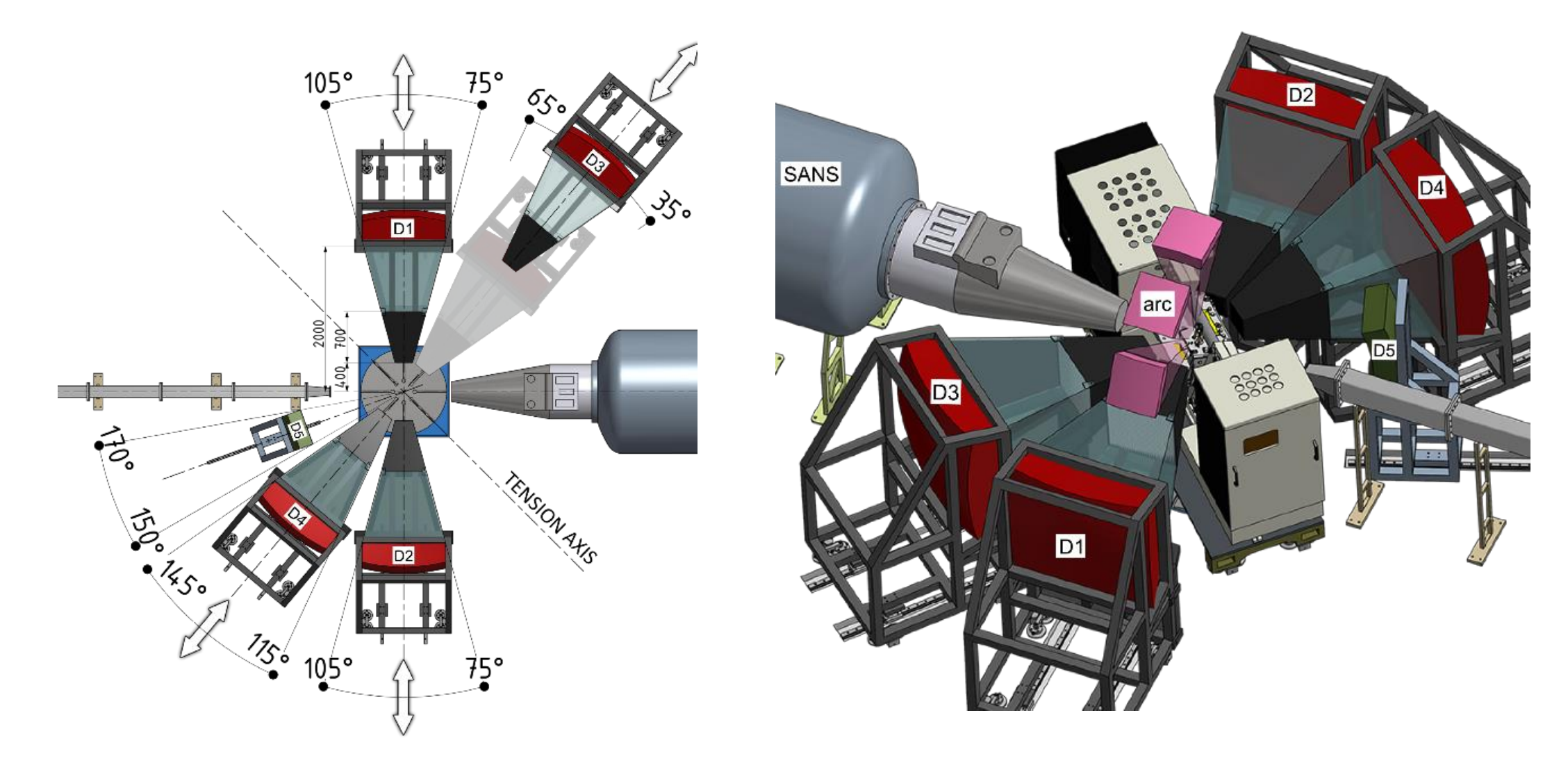}
\caption{Technical drawing of the BEER sample area showing four powder diffraction detectors (labeled as D1-D4), three flat detectors combined to an arc-shaped detector unit for texture analysis. The two grey, cuboid shaped building blocks depict the sample environment called Gleeble for sample processing. Additionally, shown at 160$^\circ$ in the left panel and labelled as D5 is a backscattering detector. Both panels also show the location of the SANS-tube where the SANS detector will be installed with its swivel-mounted collimation cone. The neutron imaging detector is not shown here, but it will be placed for measurements at a distance of 0.05 - 0.2 m from the sample, at the location of the removable SANS collimation cone. The detectors will be installed on rails that will allow them to be moved back-and-forth in order to make room for large sample environments that might be used with various samples. Taken from  \cite{beer}. }  
\label{fig: f3}
\end{figure*}

The instrument proposal foresees several detector systems for BEER in order to study the diffraction pattern and perform texture and strain analysis,  as well as SANS and NI measurements for the sample under investigation. The detectors will be positioned at 90$^\circ$, -90$^\circ$, 50$^\circ$, and -130$^\circ$, and will be mounted on rails together with the attached radial collimators, see Fig. \ref{fig: f3}. The angular span of each detector is 30$^\circ$ for a sample-to-detector distance of 2 m. Thus, the detectors will have a position resolution of $<$2$\times$5 mm$^2$ and cover $\sim$4 m$^2$ of active area (~$\sim$1 sr).  

The four powder diffraction detectors will be complemented by three arc  detectors mounted at 90$^\circ$ at a fixed distance of 1 m from the sample, see right panel of Fig. \ref{fig: f3}.  Each arc detector will have an area of circa 0.5 m x 0.5 m. A  detector with an area of 0.25 m$^2$ placed inside a vacuum tank at 6.5 m behind the sample will measure the SANS spectrum simultaneously with the high or medium resolution powder diffraction patterns. The beamline will also contain a backscattering detector that will be mounted at 160$^\circ$, see left panel of Fig. \ref{fig: f3}.   In the transmitted beam, a small position-sensitive detector with a size of $\sim$4 cm$\times$4 cm will serve to image and possibly perform energy analysis (Bragg edge), and hence map the various microstructural characteristics of the sample.  

\subsection{Detector specifications for the ESS diffractometers} 

The technical specifications for the detector systems envisaged for the three approved ESS diffraction instruments are collected in Table \ref{table:alldet}. The instrument teams intend to use diffraction detectors that follow the current trend for modern diffractometers populated with wide solid-angle 2D position-sensitive detectors in order to enable 3D data analysis. Please note that the values given in the proposals for the active area, number of detector units or even the pixel size could be subject to change, as one of the main activities foreseen  during the preliminary engineering design is to refine the instrument design and hence the detector requirements. Additional requirements for all diffraction detectors are a high detection efficiency (ca. 50$\%$ at 1.8 \AA), time resolution in the order of 10 $\mu$s, and $\gamma$-sensitivity that should not be higher than $\sim$10$^{-7}$. Obviously, all detectors types must have excellent long-term stability, uniformity and reliability.   

According to Table \ref{table:alldet}, the most severe requirements for the ESS diffraction detectors are the count rate capability and spatial resolution. The values for the time-averaged flux on the sample estimated with Monte-Carlo simulations  are up to an order of magnitude higher than those available at similar instruments operational at the newest generation of spallation sources such as TS2 at ISIS \cite{wish1}. This will result in very high instantaneous rates in detectors that will generate a large amount of data that needs to be readout and processed very quickly. Additionally,  the spatial resolution of the diffraction detectors enables studies on samples with volumes of the order of  a few mm$^3$. This requires detectors that can achieve position resolutions of 5 mm or smaller.  These issues will be discussed in more detail in the next sections.

\begin{table}
\small
\caption{Specifications of the DREAM, Heimdal, and BEER detectors as given in the instrument proposals \cite{dream,heimdal,beer}.  The last column gives the time-averaged flux on the sample in normal operation mode for SANS and NI and high-intensity (HI) mode ($\Delta$d/d$\sim$1$\%$) for powder diffraction (PD) for Heimdal and BEER and medium-intensity mode ($\Delta$d/d$\sim$0.5$\%$) for DREAM, as estimated during the Monte-Carlo optimisations of the instrument designs.  }
\label{table:alldet} 
\centering
\begin{tabular}{|c|c|c|c|c|c|c|c|} 
\hline\hline
&&&Area&&&  \\
ESS&Detector&Spatial&detector&Number&Angular&Time-averaged \\
diffraction&type&resolution&unit&of detector&coverage&flux on  sample  \\
instrument&&  (mm$^2$) &(m$^2$)& units&(or Q-range)& (n/s/cm$^2$)  \\
\hline\hline
&&&&&&\\
DREAM&PD&4$\times$4 &9.7&1&45$^\circ$-135$^\circ$&3.4$\cdot$10$^8$\\
\cite{dream}&Back-scatt.&3$\times$3&1&1&135$^\circ$-177$^\circ$&\\
&Forward&3$\times$3&1&1&1$^\circ$-45$^\circ$&\\
&&&&&&\\
&&&&&&\\
Heimdal&PD, Day-1&3$\times$10&4.5&1&10$^\circ$-170$^\circ$&2$\cdot$10$^9$\\
 \cite{heimdal}&PD, Day-2&3$\times$10 &2.5&1&55$^\circ$-150$^\circ$&\\
&SANS&4$\times$4 &1&4&0.001<$Q$<3 \AA$^{-1}$&$<$10$^6$\\
&&4$\times$4 &0.5&3&&\\
&NI&0.1$\times$0.1 &.002&1&$\sim$10 cm$^2$&\\
&Back-scatt.&3$\times$3&0.25&2&150$^\circ$-170$^\circ$&2$\cdot$10$^9$\\
&&&&&&\\
&&&&&&\\
BEER&PD&$<$2$\times$5&1&1&35$^\circ$-65$^\circ$&10$^9$\\
\cite{beer}&&&&1&75$^\circ$-105$^\circ$&\\
&&&&1&75$^\circ$-105$^\circ$&\\
&&&&1&115$^\circ$-145$^\circ$&\\
&Arc detectors&$<$2$\times$5&0.25&3&100$^\circ\times$30$^\circ$&10$^9$\\
&SANS&5$\times$5 &0.25&1&0.003<$Q$<0.15 \AA$^{-1}$&6.2$\cdot$10$^6$\\
&NI&0.05$\times$0.05 &.0016&1&$\sim$10 cm$^2$&$<$5$\cdot$10$^7$\\
&Back-scatt.&5$\times$5&0.25&1&150$^\circ$-170$^\circ$&10$^9$\\
&&&&&&\\
\hline\hline
\end{tabular}
\end{table}

\subsection{Strategy for the diffraction detectors at the ESS}

Diffraction instruments need to be able to identify the relative intensity of different Bragg reflections. Therefore the instrument performance is defined by the ability to reconstruct the intensity of these peaks and the ability to identify small features. This requires an undistorted peak height in the most intense pixels and a low background to maximise signal to noise. Thus, the detector technology of choice must be able to balance the competing requirements to measure the intensities, positions and widths of diffraction peaks simultaneously. 

The research on detectors to be employed at the European Spallation Source is an important parallel effort to the design and construction of the facility itself. These studies involve the ESS staff detector scientists and the In-Kind partners from many European countries, and  comprise of detector simulations, proof-of-concept, feasibility and validation studies,  construction of prototypes and realistic-size demonstrators, or refinements of existing designs \cite{FP1,uTPC,Jon12,Ste02,FP2}. This combined effort is needed in order to achieve the challenging high performance demanded by the frontier science cases that will be studied at the ESS, as well as a successful integration of the detectors and dedicated electronics into the future instruments.   A brief overview of the first instruments recommended in 2014 for construction at the ESS, expected detector needs, appropriate detector technology options, and a preliminary timeline for the construction schedule is given in Ref. \cite{vertex}. A firm decision concerning the detector technology and design for a specific instrument will be made during the instrument preliminary engineering design phase. The decision will be based on a detailed assessment of the science goals of the instrument, key detector requirements, maturity, reliability and maintainability of the candidate detector system and obviously, cost and schedule. 

The primary candidate detector technologies to consider for the diffraction instrument class at the ESS are those employed at similar instruments at existing facilities.  Similar requirements for each of the envisaged set-ups that will be available to the users (i.e., diffraction, SANS or NI) make it possible for a particular detector technology or design to be used by more than one instrument. For example,  at least one ESS diffractometer could exploit  scintillator-based detectors, provided this technology is able to cope with the expected high event rate and also fulfill the other requirements listed in Table \ref{table:alldet}. An excellent overview of the current status and ongoing developments of detectors utilising scintillators is given in Ref. \cite{scint}.  A detector based on the MSGC-technology and a curved-MWPC similar to those used at the D20 \cite{d20} or D1B instruments \cite{d1b}, respectively, also meet the requirements in terms of pixel size and count rate capability. However, both technologies would require a significant amount of R$\&$D and engineering work in order to increase the angular coverage in the vertical direction, which is now limited to 40 \cite{msgc} and 20 cm \cite{graham}, respectively. But, more importantly, the unclear unavailability of the $^3$He gas in the future represents a high risk for the ESS science program \cite{mpgd,rhw}. 

As shown in Table \ref{table:alldet}, the estimated time-averaged flux on the SANS samples for both HEIMDAL and BEER small-angle add-ons  is around 10$^6$ n/s/cm$^2$, which is much lower than at standalone SANS instruments \cite{d33,d22}. Moreover, the required sensitive area does not exceed 1 m$^2$ per detector. Thus, the need for high-efficiency, high-granularity SANS detectors could, in principle, be satisfied with the existing technologies such as the medium-size $^3$He-filled MWPCs similar to those operational at the REFSANS instrument at MLZ Garching  \cite{Refsans}, the SANS beamline at the BNC \cite{bnc} or the D16 instrument at ILL \cite{d16,miland},  a MSGC-based detector \cite{msgc} or even square-shaped position sensitive-tubes (PSD) in a monoblock design \cite{d33}. 

The 8-mm diameter $^3$He-filled PSDs, similar to those deployed at the WISH diffractometer at ISIS, are presently not considered for the ESS powder diffraction detectors. This type of tube, largely used at the existing neutron scattering facilities, does not provide the spatial resolution required by the ESS diffractometers, see Table \ref{table:alldet}. Thus, the ESS diffraction detectors must  employ alternative technologies. As such,  gas-filled detectors incorporating solid $^{10}$B converters  show great potential to become the replacement technology for $^3$He-tubes \cite{FP1,Jon12,Ste02,FP2,Lacy10,Hen12,Lacy14}. In a MWPC,  the $^{10}$B-coated cathodes can be mounted in two different geometries with respect to the direction of the incoming neutron: either perpendicular ($\eta=$90$^\circ$, normal geometry) or at an angle $\eta\ll$90$^\circ$ (inclined geometry).  The latter arrangement has several advantages. The inclination of the boron-layer  leads to an increase of the effective absorption film thickness by a factor proportional to $1/sin(\eta)$ and thus, to a larger detection efficiency for a single converter layer. For example, the detection efficiency for a thermal neutron striking a 3 $\mu$m thick layer at 90$^\circ$ (normal incidence) is $\sim$5 $\%$, but increases to 60$\%$ when the layer is positioned at 5$^\circ$ with respect to the direction of the incident neutron \cite{Modzel,Gregor15}.  Moreover, the wire pitch seen by the incoming neutron becomes smaller by a factor proportional to $sin(\eta)$, which at first glance improves  the position resolution and the counting rate capability of the detector by $1/sin(\eta)$ \cite{FP1}.  However, all these improvements in performance come at the cost of a more challenging engineering design and a higher manufacturing cost for the detection system.     

Experimental studies with several versions of prototypes and realistic-size detectors exploiting $^{10}$B-solid converter have been tested  side-by-side with the  $^3$He-tubes in inelastic neutron scattering experiments at ILL and SNS \cite{Lacy14,Kap15} . The results indicated that  the performance of this detector concept is close to that of the pressurised $^3$He-tube in terms of detection efficiency and neutron-$\gamma$ discrimination. The measurements also demonstrated that this detector technology is suitable for industrial production \cite{FP2,Link}. This makes boron-based detectors a viable solution for applications that require large-area neutron detectors, such as chopper spectrometers.  The typical gas mixture used to operate $^{10}$B-counters is Ar-CO$_2$, which is a cheap, commercially available, non-flammable gas with a low intrinsic sensitivity to gamma-rays. The use of pulse-height discrimination electronics to read out the signal allows for the operation of the $^{10}$B-based gas detectors at gas gains as low as 100 \cite{FP1}. This  reduces the wire-aging rate and also facilitates  the detector engineering by relaxing the mechanical tolerances.  Furthermore, the radiation hardness of the boron-thin films produced at the ESS coating facility in Link\"oping was successfully tested in recent measurements at the Prompt Gamma Activation Analysis (PGAA) instrument at the FRM-II research reactor in Garching, Germany. The tests were carried out by exposing the films to a neutron fluence of up to 1.1$\times$10$^{14}$ cm$^{-2}$, which corresponds to what the films would be irradiated with in more than 30 years of continuous operation at the ESS diffractometers  \cite{Hog12,Hog15}.  

Historically, the thermal neutron detector consisting of a gas counter lined with a thin layer of $^{10}$B-solid converter or filled with BF$_3$ gas is older than the $^3$He-based technology. The first boron-based detector was proposed in the 1930s \cite{Funf}. Boron-lined and BF$_3$-gas counters were used in applications with thermal and cold neutrons produced by nuclear reactors for several decades before $^3$He-gas  became available in large quantities at a reasonable price. The recent $^3$He-crisis generated a revival of boron-based gas detectors \cite{Kou2010,wilpert}. Nowadays this type of detector is sometimes referred to as  ``new detector technology'',  but not   in the sense of a fundamentally new invention. Owing to its low neutron detection efficiency, boron-lined tubes or variants of it were never before considered for the detection of  neutrons scattered in inelastic and diffraction measurements.  In order for the $^{10}$B-technology to become a feasible replacement for the $^3$He-gas tubes, in use at modern neutron scattering facilities, the detector system must incorporate stacks of boron layers. The engineering and operation of such a large system becomes more demanding and requires significant progress in the mechanical design,  readout electronics, data analysis, and simulation capabilities.  The experience gained in medium and high-energy physics with simulating, constructing and operating large-area position-sensitive detectors,  handling large amounts of readout channels, and developing advanced data acquisition and data reduction methods will be vital for the successful implementation of the $^{10}$B technology, although a number of issues will remain unique to the neutron scattering applications.  

At the ESS, two diffraction instrument teams pursue $^{10}$B-based detectors to record the diffraction patterns. This will make ESS the second neutron scattering facility in the world that employs this technology for applications that require large-area detectors; the first being the Neutron Source Heinz Maier-Leibnitz (FRM II) in Germany \cite{mlz}. 

The ESS detectors will be designed and built by In-Kind partners assisted by the local scientists. In the following subsections, we will give a brief description of the detector design fundamentals. The detection systems described below represent reference concepts, and therefore the technical specifications are still preliminary and subject to change. During the preliminary engineering design of the instruments, Monte-Carlo codes will be used to simulate the performances associated with different choices of the construction parameters in order to refine, and finally deliver, robust instrument designs. These analytical refinements will provide updated values for the incident flux on the sample that will also include the recent improvement in the ESS thermal moderator and optimisation of the physics output by carefully selecting the detector parameters. A complete description and in-depth details of the instruments and detectors will be made available later in dedicated publications by the respective leading teams of scientists and engineers.  
 
\subsubsection{The DREAM powder diffraction detector}\label{sec:dream_section}

The conceptual frontrunner of the DREAM powder diffractometer is the POWTEX instrument  that is currently under construction at the MLZ center in Garching \cite{mlz,powtex}. POWTEX will be a high performance time-of-flight diffractometer at a continuous source that will utilise several new concepts for beam optics, data analysis, and detection systems. The  POWTEX instrument  will  become the first powder diffractometer in the world that employs a detector technology based on solid $^{10}$B converters \cite{Mod14}.  

\begin{figure*}[ht]
\centering
\includegraphics[scale=0.75]{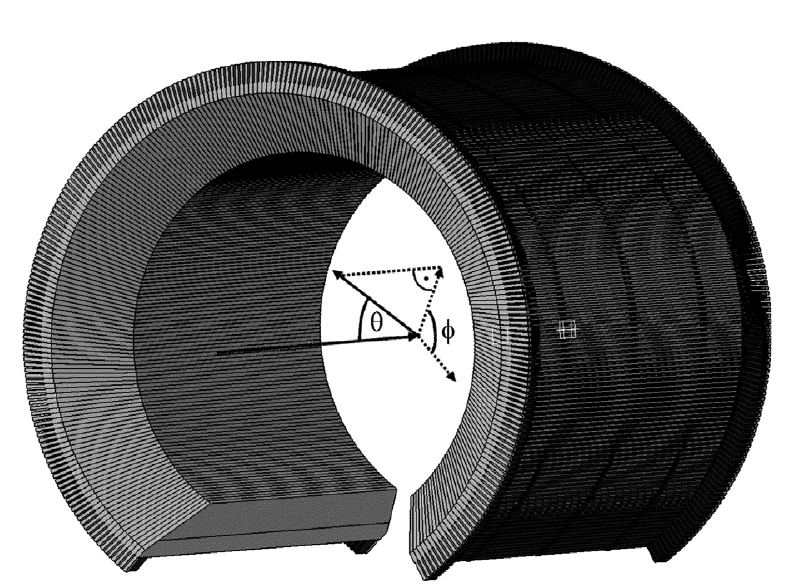}
\includegraphics[scale=0.7]{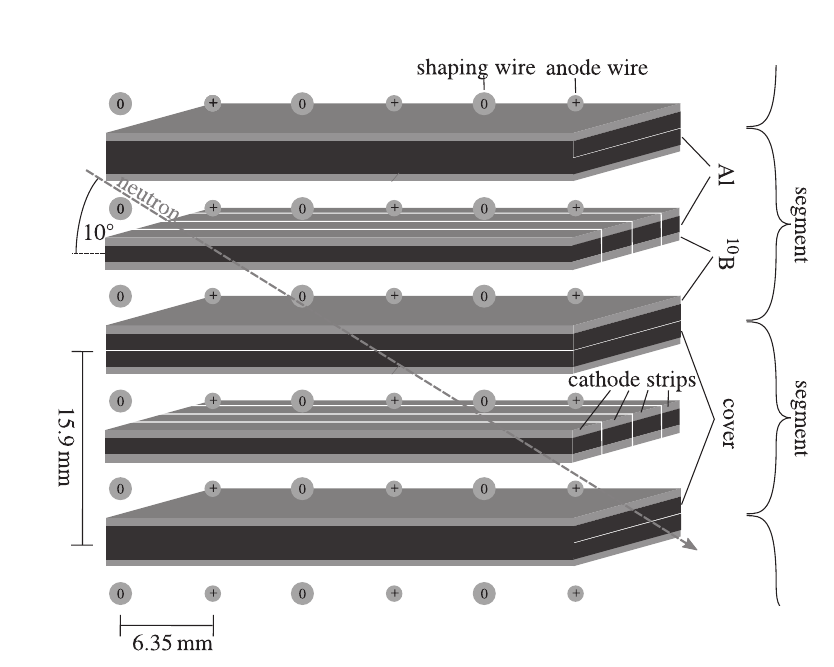}
\caption{ Left: Stacked Jalousie modules to form a large-area cylindrical detector. The drawing shows the version of the detector under construction for POWTEX@MLZ. The detector that will serve the future DREAM powder diffractometer will cover only $\sim$2$\pi$ of the area surrounding the sample, in order to allow easy access to the sample position. Taken from \cite{dream}. Right: exploded view of stacked detector segments showing the $^{10}$B-coated strips, and alternating anode and field wires.  Taken from  \cite{Mod14}.}  
\label{fig: f4}
\end{figure*}

The DREAM instrument team intends to apply  most of the new techniques developed for the POWTEX instrument, including the detector technology \cite{dream_det}. The proposed diffraction detector will be based on the $Jalousie$ design concept by CDT Heidelberg  \cite{Hen12,Mod14}. It  will consist of long, rectangular modules containing   anode wire planes and cathodes. The cathodes are made of thin lamellae arranged side-by-side and coated with a 1.2 $\mu$m thick layer of $^{10}$B.  The modules will be mounted in a cylindrical geometry around the powder sample, at an angle of 10$^\circ$ with respect to the direction of the incoming neutrons, as shown in Fig. \ref{fig: f4}.  Thus, one module with a length of  2.5 m  and a width of 0.21 m will cover a total solid angle of 2.5*0.21*sin(10$^\circ$)/(1.25$^2$)=0.058 sr.  According to calculations, a  detection efficiency of $\sim$55$\%$ at 1 \AA~ \cite{Hen12,Mod14} could be achieved by choosing the design parameters such that the incoming neutron is allowed to traverse four detection planes (four counters, i.e., eight boron-layers). The distance between the cathodes (boron layers) will be between 7 and 9 mm, while the wire pitch will be around 6 mm. All modules of the detector will  be operated  with Ar-CO$_2$ gas at atmospheric pressure. 

\subsubsection{The Heimdal diffraction detectors}
\label{sec:Heimdal_section}

The Heimdal instrument team favours the use of  scintillator technology for the neutron diffraction detectors, as PSI, one of the co-proposing partner laboratories, has already extensive expertise in the design and construction of scintillator detectors and associated electronics for neutron scattering applications \cite{poldi,poldi2,poldi3}. One of the options considered for the Heimdal powder diffraction detector is a scintillator module based on an unit already available as a prototype designed for the POLDI instrument at the SINQ laboratory in Switzerland \cite{sinq}.   Pushing the count rate capability and reducing the pixel size will be the main driving forces for the development of the Heimdal detector, which could also involve external expertise and contributions from other In-Kind partners. Another option would be to use the same detector technology that has been selected by the DREAM instrument team and discussed in the previous subsection. The HEIMDAL requirement for spatial resolution can be fulfilled by choosing the appropriate wire pitch and cathode design of the Jalousie detector module \cite{Hen12,Mod14}. 


\subsubsection{The BEER diffraction detectors}
\label{sec:beer_section}

The BEER collaboration also plan to employ $^{10}$B-based detectors to perform the strain scanning and texture analysis on engineering samples. There will be four detectors (labelled as D1-D4 in the right panel of Fig. \ref{fig: f3}), each consisting of a stack of up to 15 flat MWPCs with an area of 1 m $\times$ 1 m.  The flat geometry has the flexibility to be used at different sample-to-detector distances so as to choose either high resolution at large distances or large angular coverage at short distances. The efficiency of the stack is expected to be $\sim$60\% at 2 \AA. The required position resolution of less than 2 mm$\times$5 mm and high counting rate capability will be achieved by means of a wire pitch and an anode-cathode distance of $\sim$2-3 mm. A similar detector concept is being considered for the arc detectors. A detailed description of the BEER detector concept and design is currently being prepared \cite{hzg}.

\section{Estimation of event rates in the diffraction detectors} 
\label{sec: rate_est}

In  this section we will introduce an analytical formula to describe the dependence of the global time-averaged count rate of a diffraction detector on the time-averaged neutron flux incident on the sample under investigation.  With this formula, we aim to provide a simple analytical tool that can be easily applied whenever there is a need for a reliable estimate of the required detector rate.  A knowledge of this rate is critical in determining the appropriate detector technology, readout scheme and data handling method for a particular diffraction instrument. Without a thorough rate analysis performed before the construction of the instrument starts, there is a risk that an incorrect choice of the detector technology could be made, which will lead to an inefficient utilization of the expensive neutron beam. 

The Monte-Carlo codes commonly used in the optimisation of the instrument designs have made tremendous progress in the last few years and now include not only routines that can simulate the various instrument components, but also simulate detectors. Nevertheless  it is not wise to rely entirely on a simulation study to predict the response of a given neutron scattering instrument. The rate estimates obtained with the equation that we introduce in this section can be used in addition to those extracted from the MC-calculations, as part of the detector validation procedure. We would like to point out that the present rate analysis is not being performed with the intention of comparing the performance of the future ESS diffractometers among themselves or with that of existing similar instruments. Such an evaluation is a very complex matter and it cannot be based on the integrated flux at the sample position or the detector count rates only. The results presented in this section should be taken for nothing other than a quantitative means to assess the challenges on the detector technologies imposed by the long and bright ESS pulse. As such, four kinds of rates can be defined as relevant to the discussion here: 

\begin{itemize}
\item {\bf  Global time-averaged detector rate}, which is defined as the total number of neutrons per second recorded by the whole detector. This is relevant to designing the bandwidth in the data acquisition and storage chain. 
\item {\bf  Local time-averaged detector rate}, which is  defined as the total number of neutrons per second recorded in a detector pixel, channel or unit. The local rates for the detectors deployed at diffractometers are usually given per tube (if $^3$He-tubes are used) or PMT (if scintillator detectors are used).  For simplicity, we normalise the local rates to $cm^2$. 
\item {\bf Global instantaneous peak detector rate}, which is defined as the highest instantaneous neutron count rate on the whole detector. 
\item {\bf Local instantaneous peak detector rate},  which is defined as the highest instantaneous neutron count rate on the brightest detector pixel, channel or unit. At pulsed sources, the instantaneous rate could be more than an order of magnitude higher than the average rate as the neutron emission is concentrated in short bursts. The knowledge of this rate is important in determining whether a detector  technology is suitable to be utilised for a specific application and has impact on the design of the detector and electronics. 
\end{itemize}

Let  $\Phi_{sample}$ be the calculated, or measured, time-averaged neutron flux at the sample position. The sample has a cross-sectional area denoted by $A_{sample}$ and it is characterised by the scattering factor $s_{sample}$. Assuming that the diffraction detector covers a solid angle of $\Omega_{det}$ steradian and has an average detection efficiency  $\epsilon_{det,av}$, the global time-averaged detector rate will be given by: 

\begin{align}
\label{eq:eq_rate}
global~detector~rate~(n/s)=&~\Phi_{sample}(n/s/cm^2) \times s_{sample}\times A_{sample}(cm^2) \times \ldots \nonumber \\
&\ldots\times \epsilon_{det,av}\times \Omega_{det}(sr)/4\pi  
\end{align}

This rate equation exploits the figure-of-merit, FoM, introduced by Jorgensen {\it et al.}  \cite{Jorg}  to judge the performance of a new diffractometer design and compare it to that of existing instruments. We assume here that the background rate due to unwanted $\gamma$-events and neutrons that have not been scattered from the sample is negligible compared to the rate of the events of interest. The assumption of negligible contribution from the $\gamma$-background is not unreasonable, as values between 10$^{-5}$-10$^{-6}$  were reported for the $\gamma$-sensitivity of the detection systems operating at similar existing instruments  \cite{powgen11,gem2,syk12,wish2} and $^{10}$B-based proportional counters \cite{anton}. The ESS diffraction detectors must be able to achieve the same or better n-$\gamma$ discrimination level \cite{vertex}.  

The sample scattering factor $s_{sample}$ entering in Eq.  \ref{eq:eq_rate} represents the total probability for Bragg diffraction from all individual crystallites of a polycrystal. The probability for a Bragg reflection at an angle $\theta_B$ from the $hkl$ crystallographic plane of the polycrystal is given by \cite{with,book1}:

\begin{eqnarray}
\label{eq:eq_p}
P_{hkl}(\lambda,\theta_B)=\frac{\lambda^3}{4v_0^2}\frac{m_{hkl}F_{hkl}^2}{sin\theta_B^{hkl}}
\end{eqnarray}

where $F_{hkl}^2$ is the structure factor, which is the resultant of all waves scattered from the $hkl$ plane of the crystal, $m_{hkl}$ is the multiplicity of the $hkl$ reflection (i.e., the number of overlaps of identical structure factors) and $v_0$ is the unit cell volume. The structure factor $F_{hkl}$ depends upon both the position of each atom in the unit cell and its scattering power plus thermal vibration \cite{book1}.  The total intensity observed in a powder diffraction pattern (i.e., the sample scattering factor $s_{sample}$ in Eq. \ref{eq:eq_rate}) is the sum of all contributions from all $hkl$ reflections plus the background (which we neglect here).  Equation \ref{eq:eq_p}  indicates that the intensity  is proportional to the sum of the squares of the structure factors, which can reach values as large as $\approx$1/$e^2\approx$15$\%$. This upper value will be used in the next section to estimate  the rates in the ESS diffraction detectors.  

The  global time-averaged detector rates estimated with the help of Eq.  \ref{eq:eq_rate} and by assuming  an average detection efficiency of 50$\%$ for all ESS diffractometers, are shown in Table \ref{table:req}. The values for the integrated flux on the sample $\Phi_{sample}$, used in the calculations, are those obtained in the Monte-Carlo simulations of the respective instrument designs and made available in the proposals submitted for evaluation \cite{dream} or published elsewhere \cite{holm}. These values are likely to be revised upward by a few percent when the results of the simulations with the new moderator design described in \cite{Mezei2014} become available from the instrument teams.   In the last column of Table \ref{table:req}, the results of Eq. \ref{eq:eq_rate} are compared to the global time-averaged detector rates obtained in the Monte-Carlo simulations of the detector rates made available by the instrument teams \cite{dream,holm}.  The simulations were performed with the Na$_2$Ca$_3$Al$_2$F$_{14}$ reference sample, whose structure factors are known with good precision \cite{iucr}. 

As mentioned in the previous sections of this work, one of the most important features of the ESS diffractometers will be the flexibility to trade resolution for intensity. The simulated values for the integrated flux on the sample and global time-averaged detector rates given in Table \ref{table:req} are those obtained for the ``high-intensity'' mode of operation. This mode corresponds to the largest $\Delta d/d$-value for which the diffraction peaks can still be resolved when the instrument is optimised to deliver the highest possible flux on the sample under investigation.  Shown for comparison in the same table are the results of  a similar calculation performed using the virtual model for the WISH instrument at ISIS, also in "high-intensity" mode. This model was made using the VITESS software package \cite{vitess}.  Obviously, the quality of all the simulation results depends upon the accuracy with which the instrument geometry and the sample are described in the models.  

\begin{table}
\centering
\small
\caption{Comparison between the global time-averaged detector rates calculated with  the rate equation proposed in this section and the integrated counts in the diffraction patterns  calculated with Monte-Carlo simulation for the Na$_2$Ca$_3$Al$_2$F$_{14}$ powder sample on the DREAM and Heimdal diffractometers \cite{dream,heimdal,holm}. A similar calculation for WISH by using the time-averaged flux on the sample quoted in \cite{wish1}  is included for comparison.  A scattering factor of 5$\%$ was used in the calculations performed with equation 3.1, as determined from the VITESS simulation for the Na$_2$Ca$_3$Al$_2$F$_{14}$  compound \cite{vitess}.}
\label{table:req} 
\begin{tabular}{|c|c|c|c|c|c|c|c|c|} 
\hline\hline
&&&&&Global time-averaged&Global time-averaged\\
&Area&Time-averaged&&Area&rate in detector,&rate in detector,\\
Instrument&detector & flux on sample&$\Delta$d/d@90$^\circ$&sample&\bf {Eq. \ref{eq:eq_rate}}&Monte-Carlo\\
& (sr)&(n/s/cm$^2$) &($\%$)&(cm$^2$)&(n/s)&(n/s)\\
\hline\hline
DREAM&6.2&3.4$\cdot$10$^8$&0.5&0.8&3$\cdot$10$^6$&10$^7$\\
(ESS)&&&&&& \cite{dream}\\
&&&&&&\\
Heimdal&2.25&1.5$\cdot$10$^9$&1&1&6$\cdot$10$^6$&5$\cdot$10$^6$ \\
(ESS)&&&&&&\cite{holm}\\
&&&&&&\\
WISH&1.21&1.1$\cdot$10$^8 $&0.5&1&3$\cdot$10$^5$&4$\cdot$10$^5$\\
(ISIS)&&&&&& [this work]\\
&&&&&&\\
\hline\hline
\end{tabular}
\end{table}

As shown in Table \ref{table:req}, the results of  our analytical calculations are in close agreement with the integrated number of counts in the diffraction spectra obtained from the  Monte-Carlo simulations.  The results of both methods indicate that the global time-averaged event rates in the ESS diffraction detectors  can be as large as 10 MHz, which is over an order of magnitude larger than the event rate estimated for the WISH instrument in ``high-intensity'' mode. However, an important contribution to the global time-averaged  event rate arises from the angular coverage of the detector system, which for the diffractometers  included in Table \ref{table:req}, can differ by a factor five. As equation \ref{eq:eq_rate} contains the explicit dependence of the event rate on the solid angle of the detector, we can easily apply it to determine the local time-averaged detector rate by dividing the global time-averaged detector rate by the total detector area. This  might seem incorrect at first glance as the intensity of the diffracted beam can vary greatly from detector to detector (or from pixel to pixel). However, in the powder diffraction measurements performed at a pulsed source, the scattered intensity is much more uniformly distributed among the detectors than in a similar measurement at a reactor instrument. For monochromatic powder diffraction, the detectors (or the detector pixels) located at the $\theta_B$ angles that satisfy the Bragg relation $\lambda$=2$dsin\theta_B$ will collect most of the scattered intensity, and this can lead to large differences in the count rates recorded across the entire detector area. In TOF-diffraction, the incident beam contains multiple wavelengths. Therefore, Bragg reflections are observed at all scattering angles, of course each with different wavelengths, which, as it will be shown below with real measurements, results in a much more uniform distribution of the time-averaged count rates in detectors.     

As mentioned above, at spallation sources the neutron emission is concentrated in short bursts, therefore  the parameter that ultimately determines the instrument count rate requirement is the local instantaneous peak neutron flux. Some of the time bins within one beam frame  can record a much larger scattered intensity and therefore, can give rise to local instantaneous peak rates that exceed the time-averaged detector rates by an order of magnitude or more (the bin width is usually of the order of tens of  $\mu$s. One beam frame $\approx$ 1/source repetition rate, i.e., $\approx$100 ms for WISH \cite{wish1} and $\approx$20 ms for GEM \cite{gem1}). Thus, the instrument teams must select detector technologies that have a wide dynamic range and are able to cope with the maximum intensity delivered during a milliseconds-long time frame. In the reminder of this subsection we present a simple way to estimate the local peak detector rates, which is based on a comparison with real experimental data obtained from existing similar instruments. 

\begin{figure}[tb]
\centering
\includegraphics[scale=0.4]{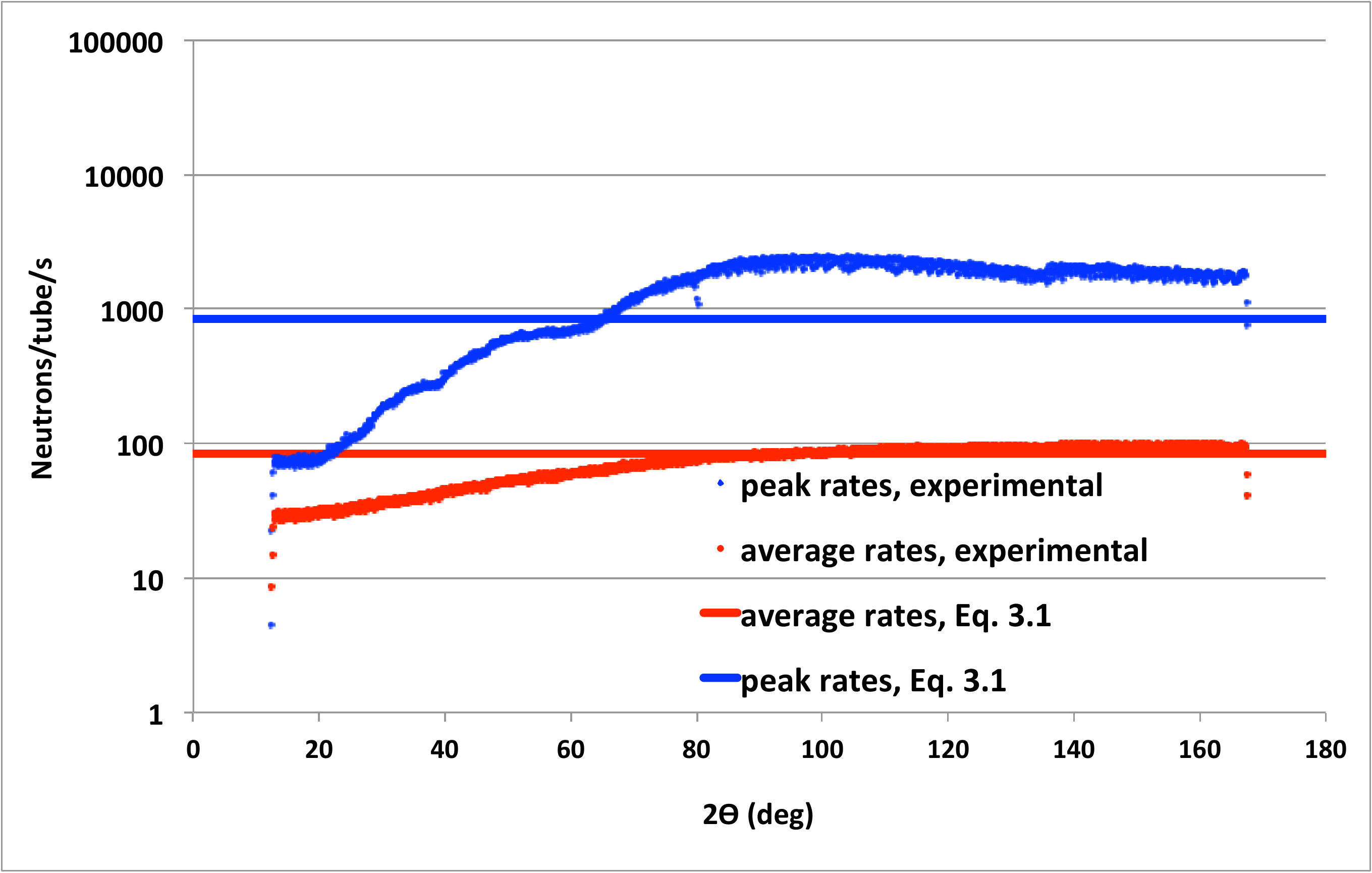}\\
\includegraphics[scale=0.4]{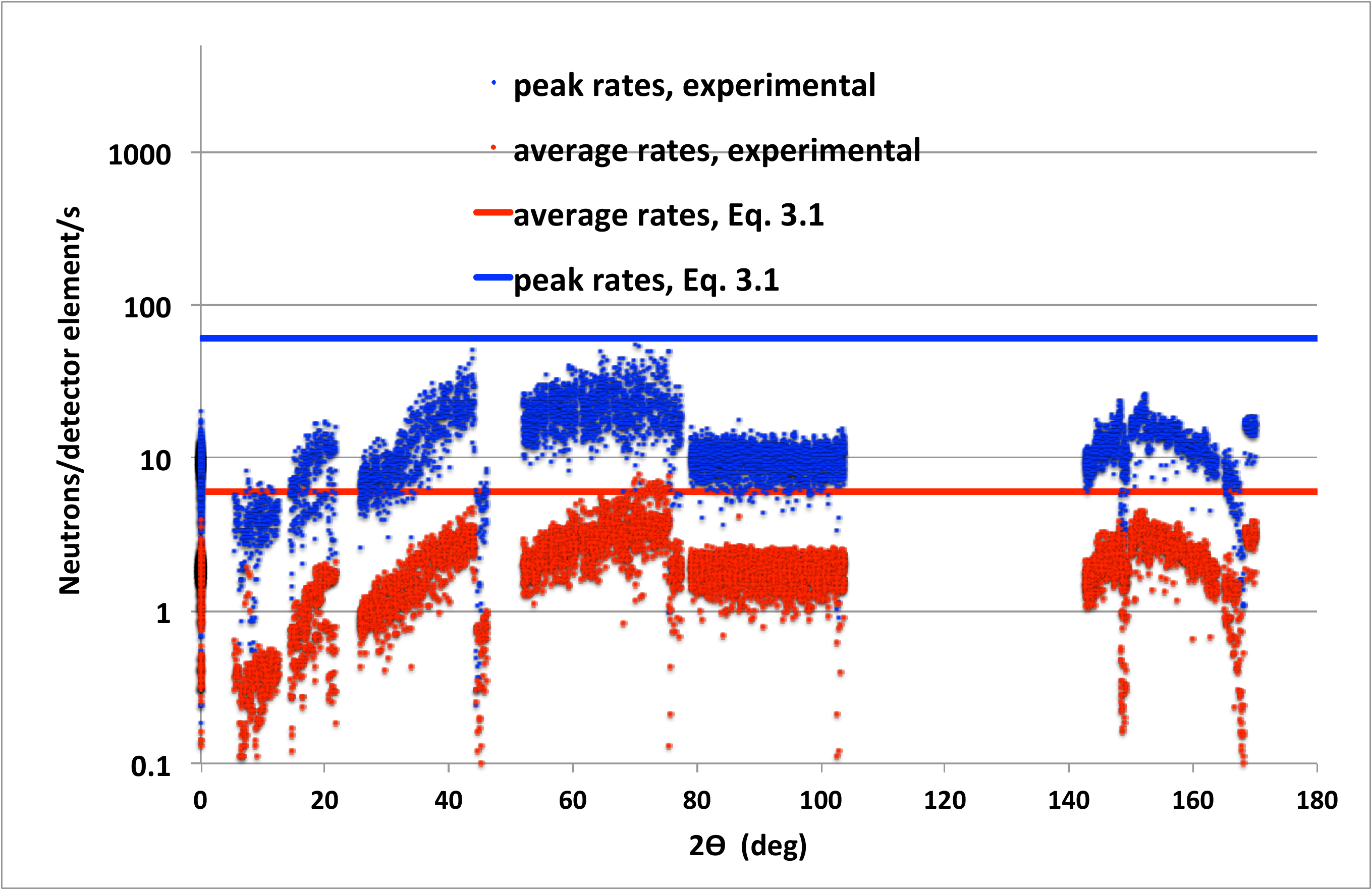}
\caption{Distribution of the neutron events per second detected in the WISH position-sensitive detectors (top) and GEM scintillator elements (bottom) as a function of the scattering angle 2$\theta$. The experimental data were collected with the  same Na$_2$Ca$_3$Al$_2$F$_{14}$ reference sample. The experimental error bars are smaller than the size of the symbols. The red horizontal lines correspond to the rates estimated with Eq. 3.1, by using the time-averaged flux values quoted in literature for both instruments \cite{wish1,d20} and a sample scattering factor of 5$\%$, as extracted from the VITESS simulation with the Na$_2$Ca$_3$Al$_2$F$_{14}$ sample \cite{vitess}.}  
\label{fig:f6}
\end{figure}

The data sets used in this work to provide  empirical support to our rate analysis were diffraction patterns collected with the same Na$_2$Ca$_3$Al$_2$F$_{14}$ powder sample studied with both the WISH and GEM diffractometers operational at the ISIS facility in the UK \cite{isis}.  This sample is representative for materials within the scope of powder methods as its diffraction pattern consists of around 300 reflections in the $d$-range between 0.7 and 4 \AA~\cite{iucr}, the most significant range to explore polycrystalline powders and engineering materials (i.e., materials that contain large amounts of Fe, Al, Ni, Cu or Ti). The two diffraction instruments selected for this study are at the forefront of their class and feature complementary detector technologies that exploit a different architecture around the sample position. While WISH makes use of identical $^3$He-tubes arranged in a cylindrical geometry at constant distance from the sample \cite{wish1}, GEM operates a 9-bank detection system, each bank consisting of several ZnS/$^6$LiF scintillator modules mounted in a {\it resolution-focused geometry}, i.e., all elements within each bank are arranged to have an approximately constant resolution \cite{gem1}.

The WISH diffractometer is equipped with a substantial detector array consisting of 10 panels, each supporting 152 vertically-oriented position-sensitive tubes filled with 15 bar of $^3$He \cite{wish1}. The whole detector covers the scattering angles between 10$^\circ$ and 170$^\circ$. The tubes are 1 m long and have a diameter of 8 mm. The length of a tube is electronically divided into 128 channels, each having a nominal resolution of 8 mm, which matches the horizontal spatial resolution \cite{wish1,wish2}. The detection efficiency of a tube is 50$\%$ at 1 \AA. A GEM detector element has a typical active area of 5$\times$200 mm$^2$ and is composed of two strips of ZnS/$^6$LiF scintillator material arranged in a V-shape coupled to photomultipliers via fiber optic cables \cite{gem2}. Typically circa 120 of such detector elements are grouped together in modules arranged such that each element is tangential to a Debye-Scherrer cone of diffraction, with the two ends of the element equidistant from the sample \cite{gem1}.  The modules are organised in 9 banks, with all modules in one bank having the same design. The reported efficiency of a module is 50$\%$ at 1 \AA~\cite{gem2}. 

The recorded data  were stored in  {\it nxs}-file format (i.e., Nexus format \cite{nexus}), which contained the number of neutron events per detector pixel (element) and per time frame  \cite{mantid}. The average rates were determined by dividing the sum of the neutron events recorded in all available time-of-flight  bins by the duration of the measurement. The peak rates are given by the content of the TOF-bin with the highest observed signal divided by the width of the time bin and the number of time frames. The quality of the Na$_2$Ca$_3$Al$_2$F$_{14}$ diffraction patterns collected with the WISH detector array can be assessed by inspecting the spectra published in Ref. \cite{wish2}.  

The experimental average and peak rates are displayed in Fig. \ref{fig:f6}  as a function of  the scattering angle $2\theta$, which corresponds to the position of the detector element in the system. For both instruments, the average count rates, represented with red symbols in Fig. \ref{fig:f6}, vary only smoothly across the WISH detectors and the elements of a GEM bank.   This observation supports the assumption made above that in TOF-measurements, the time-averaged intensity of the diffracted beam is distributed rather uniformly among the detectors.  
Moreover, the experimental peak rates, corresponding to the maximum of the most intense peak collected  during one time frame and represented with blue symbols in the same figure, are higher than the time-averaged rates, as the latter are determined by averaging out over all time frames. As a general trend, the higher the density of the Bragg peaks, as it is the case of polycrystalline materials such as Na$_2$Ca$_3$Al$_2$F$_{14}$, the larger the average rate in detectors and the lower the ratio between the peak and the average rates. This is also the case for non-crystalline diffraction on samples made of disordered materials, where the scattered intensity is distributed relatively evenly in the $Q$-space, rather than being concentrated in large Bragg peaks \cite{gem1}.

In standard operation, WISH provides data over the $d$-spacing range from 0.7 to 10.1 \AA~with a single frame bandwidth of $\sim$9.4 \AA~and a peak flux at 2.8 \AA~\cite{wish1}. The large WISH bandwidth allows to measure large slices of the $d$-spacing domain in all detector panels. For example, the first panel covers the angles between 10 and 42$^\circ$ and it yields data for $d$-spacings between 0.98 and 56.8 \AA. However, the reflections observed from the range with low-$d$ spacings, although characterised by large scattering factors, have low statistics. This is due to the decreasing  neutron flux when moving away from the 2.8 \AA~peak-value down to lower wavelengths, a feature of the solid-methane moderator viewed by WISH \cite{wish1}. In general, the data collected from the first two detector panels are used to investigate materials with large unit cells, which give useful information at large $d$-spacings. The peak resolution is larger but this is not so critical given the usual lower density of Bragg peaks in this region. The detector panels located at angles around 90$^\circ$ and higher collect the reflections from the most significant $d$-spacing domain for the Na$_2$Ca$_3$Al$_2$F$_{14}$ sample at the peak incident flux and with improved resolution. The diffraction data observed with  these detectors are dominated by the peaks at 1.8132 and 2.1868 \AA, which have the  largest spectral weight \cite{iucr}. This is reflected in the rapid increase of the peak detector rates as opposed to the smooth rise of the average rate, see top panel of Fig. \ref{fig:f6}. 

The lower and upper limits of the GEM wavelength spectrum for the incoming neutrons are 0.5 \AA~and 4 \AA, respectively, peaking at around 2  \AA~\cite{gem1}. As this instrument is viewing the ISIS liquid-methane moderator, the flux at short-wavelengths ($<$ 1 \AA) is larger than that of WISH.  This is advantageous for the investigation of samples that feature the main reflections at low and medium $d$-spacings. The peak and average detector rates  observed with this instrument follow the same trend with increasing the scattering angle at an almost constant ratio ($\sim$10), see the lower panel of Fig. \ref{fig:f6}. Banks 3, 4 and 5 have the largest number of detector elements and cover the largest angular ranges, therefore these banks  capture the most important part of the diffraction pattern of the sample, although with different resolutions. As each detector bank is located at a different distance from the sample, the solid angle covered by the modules installed in the different banks is different \cite{gem1}. The greatest solid angle is available around 65$^\circ$ (bank 4), which might explain the slightly larger count rates observed with this detector bank. The most backward detectors allow to explore the $d$-spacings features in the range  0.25-2.11 \AA, which is very narrow compared to the most backward WISH panel. The lack of sufficient cold-neutron flux from the liquid-methane moderator prevents one from employing the GEM banks 6 and 7 to observe the reflections from the medium and long $d$-spacings in the Na$_2$Ca$_3$Al$_2$F$_{14}$ sample. Thus, the absence of reflections in this range, including the most intense Bragg peak at 2.1868 \AA~\cite{iucr}, leads to a decrease of the data collection rate at large angles, see low panel of Fig. \ref{fig:f6}.    

The experimental data presented above suggests an order of magnitude difference between the peak and the average detector count rates measured with the Na$_2$Ca$_3$Al$_2$F$_{14}$ sample investigated with two diffractometers that are quite different in terms of pulse repetition rates and pulse duration, moderator type, peak flux on sample, as well as detector arrangement. Both the peak and average rates (in absolute value) depend on these factors, but most of these dependencies are eliminated when taking their ratio. The observed data suggests that the variation of the peak and average rates with the scattering angle is determined by the details of the  sample under investigation (peak positions, density and intensities) and the flux profile of the incident beam.  The peak-to-average ratio reaches a maximum when the peak flux of the incident beam matches a large slice of the $d$-spacing domain representative for the sample under investigation, which usually occurs in backscattering.  

At the ESS, the HEIMDAL instrument  will operate with a narrow bandwidth of $\Delta\lambda$=1.7 \AA~coinciding with the maximum brightness from the thermal moderator \cite{heimdal,heimdal_nima}. BEER and DREAM will use bi-spectral extraction in order to benefit from both thermal and cold beams from the source \cite{dream,heimdal,beer,heimdal_nima}. The standard setting for DREAM covers a frame 0.5 \AA$<\lambda<$4.3 \AA, but the chopper systems allows for shifting to larger wavelengths \cite{dream}. The natural bandwidth of BEER is $\Delta\lambda$=1.73 \AA, but it can be extended via pulse suppression to 2 or 3$\cdot\Delta\lambda$ \cite{beer}. A rather constant peak flux throughout the selected bandwidth is a general requirement for all diffraction instruments, as each $\lambda$-value within must provide roughly equally useful information \cite{mezei1}. Obviously, selecting only a small interval from the long pulse in order to tune the $\Delta\lambda/\lambda$ resolution in the range suitable for diffraction partly cancels the advantage of very high integrated peak brilliance from the ESS source. Thus, the highest expected value for the time-average flux on sample is around 1.4$\cdot$10$^9$ n/s/cm$^2$, which was calculated for the HEIMDAL instrument in ``high-intensity'' mode (corresponding to $\Delta d/d$=0.66$\%$ at 90$^\circ$),  see \cite{heimdal_nima}. This value for the flux is ``only'' one order of magnitude larger than the flux achieved when WISH operates in ``high-intensity'' mode \cite{wish1}.  This immediately suggests that the absolute values for the average and peak rates expected in the ESS diffraction detectors will be up to a factor ten larger than those observed at WISH with the same sample and if the same detector geometry (and efficiency) is assumed for all instruments. However, the dependence of these rates on the scattering angle will still be slightly different owing to the different shapes of the $\Phi(\lambda)$-distribution of the incident neutron beam, which is dictated by the moderator type viewed by each instrument. 

Thus, we can conclude that for a large number of measurements including polycrystalline powders and technology-relevant materials one can expect that the peak-to-average ratio determined for the ISIS diffractometers is a good approximation for the ESS diffraction detectors. The predictions for the average detector rates calculated with Eq. \ref{eq:eq_rate} are represented in Fig. \ref{fig:f6} with red horizontal lines. During the collection of the Na$_2$Ca$_3$Al$_2$F$_{14}$ data sets, the slits were set to the values that correspond to the standard GEM operation \cite{gem1} and the operation in high resolution, low intensity mode for WISH \cite{wish1}. Thus, the values for the time-averaged flux of the sample used in Eq. \ref{eq:eq_rate}  were 1.08$\cdot$10$^7$ n/s/cm$^2$ for WISH \cite{wish1} and 2$\cdot$10$^6$ n/s/cm$^2$ for GEM \cite{d20}.  As seen in the figure, equation \ref{eq:eq_rate} provides a good estimate for the experimental time-averaged rates measured with both instruments. Obviously, the analytical formula does not account for the observed variation of the count rate with the scattering angle. The blue horizontal lines correspond to the estimated instantaneous peak rates and were obtained by multiplying the time-averaged rates obtained with Eq. \ref{eq:eq_rate} by the factor 10 suggested by the experimental observation.  

\section{Rate estimates for the ESS diffraction detectors} 
\label{sec: rate_req}

The significant factors that define the rate capability of a detection system are its ability to record the neutron-generated signals fast and the readout strategy. For the two detector technologies discussed here, gas-counters and scintillator-based detectors, the limitations in the count rate capability arise from the pulse length and the scintillator afterglow, respectively.  Appropriate signal processing strategies must also be considered in order to reduce the volume of the data transmitted by the data acquisition system and committed to mass storage units. 

Assuming that the sample to be investigated by the future ESS diffractometers will have a scattering factor of 15$\%$,  Eq. \ref{eq:eq_rate} suggests local time-averaged event rates of  $\sim$110, 400 and 60 Hz/cm$^2$ for the DREAM, Heimdal and BEER diffraction detectors, respectively. These results are shown in Table \ref{table:rate_exp}. If we further assume an order of magnitude difference between the local time-averaged and the local instantaneous peak rates, see Fig. \ref{fig:f6}, the selected detector technologies  must be able to cope with count rates as high as 4 kHz/cm$^2$; see the estimates for Heimdal in Table \ref{table:rate_exp}. Such a rate is well within the capability of both the wavelength-shifting-fiber scintillator and gas-counter technologies. 

It is well known that a MWPC used in high rate experiments can suffer from signal pileup and a significant build up of space charge by the positive ions created near the wire during the amplification process. This leads to a decrease in detection efficiency and a degradation of the position resolution. The signal pileup can be reduced by using fast individual wire readout electronics \cite{miland,detni}, and the space charge effects can be minimised by careful selection of the geometry and working parameters of the wire counter. A small gas gap (given by the anode-to-cathode distance) and a fine wire pitch ensure high speed and reduce the space charge effects expected at high counting rates. These aspects were taken into account  when selecting the wire-counter detector technology for the powder diffraction detectors for the future ESS diffractometers. However, care needs to be taken to ensure that the anode cathode gap remains sufficiently large such that detector stability meets requirements. The detector design selected by the DREAM instrument team will feature wire planes mounted at 10$^\circ$ with respect to the direction of the incoming neutrons. In such a geometry,  the wire pitch seen by the incoming beam of neutrons will be 6.6 mm $\cdot\sin(10^\circ)$=1.14 mm. This will help to spread out the beam spot over several wires, and thus increase the count rate capability of the detector.  The BEER instrument team plans to employ the same detector technology as DREAM, but with the scattered neutron beam impinging on the detector surface at normal incidence angle, with the wires mounted at 2 mm and the anode-cathode distance also set to 2-3 mm \cite{hzg}. The compact geometry of the gas counter and its operation at atmospheric pressure decrease the collection time of the ions, which enhances the counting performance of the detector.  

Systematic measurements of the rate capability of a MWPC with a geometry similar to that proposed for the BEER diffractometer and operated in various Ar-CO$_2$ gas mixtures indicate that the gas gain starts to drop at particle rates above 10$^3$ kHz/cm$^2$ \cite{And}. This is almost 3 orders of magnitude above our estimated peak event rates shown in Table \ref{table:rate_exp}. This margin is comfortably large  to ensures that the performance of the ESS instruments that will use wire-chambers is not significantly affected by the count rate capability of the chosen detector technology.  More quantitative statements concerning the performance of the diffraction detectors for ESS will be derived from further analysis of existing data and refinement of the instrument and detector designs, as well as from measurements with realistic-size detector prototypes, which are foreseen in the course of the next years.  

\begin{table}
\centering
\small
\caption{Event rate estimates  for the diffraction detectors foreseen for the DREAM, Heimdal and BEER instruments at the ESS, for powder samples with a 15$\%$ scattering factor. The detector technology that is proposed by the respective instrument teams is also shown. }

\label{table:rate_exp} 
\begin{tabular}{|c|c|c|c|} 
\hline\hline
Diffraction&Local time-averaged rate&Local instantaneous peak rate&Detector technology\\
instrument&\bf{Eq. \ref{eq:eq_rate}}&(\bf{Eq. \ref{eq:eq_rate}} X 10)&proposed \\
&(Hz/cm$^2$)&(Hz/cm$^2$)&by the instrument team\\
\hline\hline
&&&\\
DREAM&$\sim$110&1100&MWPC with $^{10}$B-coated \\
&&& cathodes at 10$^\circ$\\
&&&\\
Heimdal&$\sim$400&4000&ZnS/$^6$LiF-based scintillators\\
&&&or\\
&&&MWPC with $^{10}$B-coated\\
&&& cathodes at 10$^\circ$\\
&&&\\
BEER&$\sim$60&600&MWPC with $^{10}$B-coated\\
&&& cathodes at 90$^\circ$\\
&&&\\
\hline\hline
\end{tabular}
\end{table}

We would like to point out that the peak rates given in Table \ref{table:rate_exp} represent as far as possible upper limits as can be given for samples that exhibit reflections in the $d$-range from 0.5 to 4  \AA, which is the main range of interest for diffraction experiments.  This covers a large number of polycrystalline powders and materials of technological importance, for the investigation of which a large part of the beam time is allocated at the existing diffractometers. This is also expected to be the case at the ESS. 

An important part of the beam time at the future  ESS diffractometers will also be allocated to investigations that cannot be done elsewhere due to the low intensity of the beam, insufficient instrument resolution or detector coverage (e.g.,  measurements on very small samples ($<$ 1mm$^3$), fast irreversible kinetics, measurements of very weak scattering phenomena on reasonable timescales, etc. \cite{tdr}). Such measurements are generally not expected to challenge the detectors in terms of count rate capability. However, as powder diffraction is highly interdisciplinary, there will a number of new materials with complex structure from various branches of science that could  lead to large scattered intensities following e.g., a chemical reaction, heat treatment or cooling of the sample. The detector rates are difficult to predict for such cases.  

Last but not least, it should be mentioned that the present study does not cover the case of diffraction on single-crystals. Such measurements can be considered as the extreme case of a high count rate scenario,  in which the peak detector rates could be several order of magnitude larger than the time-averaged rates. The  analysis of the count rates expected in single-crystal diffraction measurements requires a different approach, owing to the particularities of the diffraction pattern (low peak density, highly localised reflections) for which most of the assumptions made in the present work do not apply. 

\section{ Conclusions}

Several of the instruments included in the ESS instrument suite as specified in the Technical Design Report published in 2013 will soon be entering the preliminary engineering design phase. As such, DREAM, Heimdal and BEER represent the core of the ESS diffraction suite, covering the widest possible user community and several key scientific areas.  These instruments seek to exploit the full capability of the ESS source and aim to set new standards in neutron diffraction and become the world-leading in their class. The construction of these instruments will push the limits of current technologies in several areas, such as neutron optics, chopper systems, detectors, readout electronics and data handling. 

The scientific goals and operating conditions at the ESS impose very high requirements on the position resolution and readout speed of the detectors. Pressurised $^3$He-tubes are presently not considered for use at powder diffractometers, owing to the limitation to circa 8 mm in the diameter of the standard, commercial tube, the large areas required and unclear availability of the  $^3$He gas in the future.  The technology selected by the instrument teams for two of the future ESS diffractometers is based on stacked MWPC with $^{10}$B-coated cathodes, which seems to provide solutions to several of the demands of the new generation of  diffraction instruments. This technology is in a sufficiently advanced stage now to start relating the design specifications of future demonstrators to the particular requirements of the diffraction class of instruments in a quantitative way. A detector based on scintillator technology is currently under consideration for the Heimdal diffractometer, which is expected to deliver the highest  incident flux ever achieved on a powder sample. 

It is expected that the detailed conceptual design for the powder diffraction detectors for the ESS will become available at the end of the preliminary engineering design phase of the respective instruments. The emerging  final designs must not only meet the requirements set by the scientific goals, but also find the optimum balance between cost and performance, and ensure that all components can be realised and delivered on time. 

In this work we also propose an analytical formula that can be applied to estimate the local time-averaged event rates in the diffraction detectors. The proposed equation relates the global time-averaged detector rate to the time-averaged flux on the sample estimated with Monte-Carlo simulations, the desired detector angular coverage and efficiency and the sample scattering factor. However, as at spallation sources the neutron beam  is delivered in short bursts, the instantaneous peak event rate is the relevant indicator for the required level of performance for the detector technology in terms of count rate capability.  We used real data recorded with existing similar instruments featuring state-of-the-art detection systems to determine the extent of the difference between the average and peak rates, and used that factor in our rate analysis for the ESS diffraction detectors. A reliable estimate of the expected detector rates is crucial at this stage of the ESS project, when several important decisions need to be made concerning the detector technology, readout electronics, data analysis and data handling methods. The analytical formula introduced here and the MC-simulation of a diffraction instrument that also includes the detector response, provide independent results which can be compared and cross-checked against each other. This gives us a triangle of tools, analytical - Monte-Carlo - data extrapolation, that allows an informed decision to be taken to determine the most appropriate detector technology. The present detector rate analysis  addresses the powder diffraction detectors, but the proposed formalism could be easily adapted to other instrument classes.  The results indicate that $^{10}$B-based proportional wire chambers and scintillator-based detectors, the potential technologies for the three diffractometers discussed here,  will be able to cope with the increased count rate requirements imposed by high beam intensity from the ESS source. 

Along with the improvements in detector performance it is of paramount importance to develop fast readout electronics that is able to cope with the predicted peak detector rates. The quest for finer position resolution comes at the expense of a higher number of pixels per detector module. This has the advantage that it reduces the rate per channel, thereby improving the rate capability of the detector, but also leads to an increase in the number of electronics channels and associated cost. The future detection systems will benefit directly from contributions from the In-Kind partners  who will work closely together with the local ESS staff on the design and delivery of the detectors and associated electronics and the planing of the instrument commissioning.

\acknowledgments

This work was partially funded by EU Horizon2020 framework, BrightnESS project 676548.

\end{document}